\documentclass[preprint,showpacs,preprintnumbers,amsmath,amssymb]{revtex4}

\usepackage{graphicx}
\usepackage{dcolumn}
\usepackage{bm}

\newcommand{\qed}{\nobreak \ifvmode \relax \else
      \ifdim\lastskip<1.5em \hskip-\lastskip
      \hskip1.5em plus0em minus0.5em \fi \nobreak
      \vrule height0.75em width0.5em depth0.25em\fi}

\begin{document}

\preprint{}

\title{Two-Qubit Separability Probabilities: A Concise Formula}
\author{Paul B. Slater}%
\email{slater@kitp.ucsb.edu}
\affiliation{%
University of California, Santa Barbara, CA 93106-4030\\
}%
\date{\today}

\begin{abstract}
We report a concise answer--in the case of $2 \times 2$ systems--to the fundamental quantum-information-theoretic question as to "the volume of separable states" posed by {\.Z}yczkowski, Horodecki, Sanpera and Lewenstein ({\it Phys. Rev. A}, {\bf 58}, 883 [1998]). We proceed by applying the Mathematica command FindSequenceFunction to a series  of  conjectured Hilbert-Schmidt generic $2 \times 2$ (rational-valued) separability probabilities $\pi(\alpha)$, $\alpha = 1, 2,\ldots,32$, with $\alpha=1$ indexing standard two-qubit systems, and $\alpha=2$, two-quater(nionic)bit systems. These 32 inputted values of $\pi(\alpha)$--as well as 32 companion non-inputted values for the half-integers,  
$\alpha = \frac{1}{2}$ (two-re[al]bit systems), $\frac{3}{2},\ldots \frac{63}{2}$--are advanced on the basis of high-precision probability-distribution-reconstruction computations, employing 7,501 determinantal moments of partially transposed $4 \times 4$ density matrices. The function 
$P(\alpha)$ given by application of the command fully reproduces {\it both} of these 32-length sequences, and an equivalent outcome is obtained if the half-integral series is the one inputted. The lengthy expression (containing six hypergeometric functions) obtained for $P(\alpha$) is, then, impressively condensed (by Qing-Hu Hou and colleagues), using Zeilberger's algorithm. For generic (9-dimensional) two-rebit systems, $P(\frac{1}{2}) = \frac{29}{64}$, (15-dimensional) two-qubit systems,  $P(1) = \frac{8}{33}$, (27-dimensional) two-quaterbit systems,  $P(2)=\frac{26}{323}$, while for generic classical (3-dimensional) systems, $P(0)=1$. 
\end{abstract}


\maketitle

\begin{figure}
\includegraphics{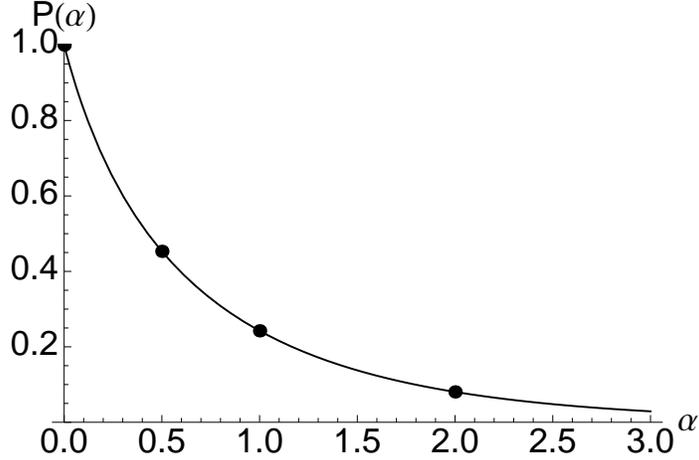}
\caption{\label{fig:HouGraph}Generalized two-qubit separability probability function $P(\alpha)$, with $P(0) =1, P(\frac{1}{2}) =\frac{29}{64}, P(1)=\frac{8}{33}, P(2)= \frac{26}{323}$ for generic classical four-level ($\alpha =0$), two-rebit ($\alpha =\frac{1}{2}$), two-qubit ($\alpha =1$) and two-quaterbit ($\alpha =2$) systems, respectively.}
\end{figure}

We present a succinct formula $P(\alpha)$ of fundamental quantum-information-theoretic importance. It effectively addresses, as well as considerably expands upon, the 
$2 \times 2$ instance of the question posed by {\.Z}yczkowski, Horodecki, Sanpera and Lewenstein in their highly-cited 1998 paper, "Volume of the Set of Separable States"  \cite{ZHSL}. These authors gave "three main reasons of importance"--philosophical, practical and physical---for studying the question of "how many entangled or, respectively, separable states there are in the set of all quantum states". The concise formula $P(\alpha)$ takes the form (Fig.~\ref{fig:HouGraph})
\begin{equation} \label{Hou1}
P(\alpha) =\Sigma_{i=0}^\infty f(\alpha+i),
\end{equation}
where
\begin{equation} \label{Hou2}
f(\alpha) = P(\alpha)-P(\alpha +1) = \frac{ q(\alpha) 2^{-4 \alpha -6} \Gamma{(3 \alpha +\frac{5}{2})} \Gamma{(5 \alpha +2})}{3 \Gamma{(\alpha +1)} \Gamma{(2 \alpha +3)} 
\Gamma{(5 \alpha +\frac{13}{2})}},
\end{equation}
and
\begin{equation} \label{Hou3}
q(\alpha) = 185000 \alpha ^5+779750 \alpha ^4+1289125 \alpha ^3+1042015 \alpha ^2+410694 \alpha +63000 = 
\end{equation}
\begin{displaymath}
\alpha  (5 \alpha  (25 \alpha  (2 \alpha  (740 \alpha
   +3119)+10313)+208403)+410694)+63000.
\end{displaymath}

Now, we claim that  the specific values of interest given by $P(\frac{1}{2}) = \frac{29}{64}, P(1) = \frac{8}{33}$, and $P(2)=\frac{26}{323}$ are the separability probabilities (with respect to Hilbert-Schmidt measure \cite{szHS,ingemarkarol}) of generic (9-dimensional) two-rebit systems, (15-dimensional) two-qubit,  and  (27-dimensional) two-quater(nionic)bit systems, respectively (cf. \cite{batle2}). Further, $P(0) =1$, for generic (3-dimensional) classical systems.

The derivation of these results--computationally/numerically strongly convincing, though still not formally rigorous in nature--was begun in the paper "Moment-Based Evidence for Simple Rational-Valued Hilbert-Schmidt Generic $2 \times 2$ Separability Probabilities" \cite{MomentBased}.  Two basic sets of analyses were conducted there. 

The first  set was concerned with establishing formulas for the bivariate determinantal product moments 
$\left\langle \left\vert \rho^{PT}\right\vert ^{n}\left\vert \rho\right\vert
^{k}\right\rangle ,k,n=0,1,2,3,\ldots,$ with respect to Hilbert-Schmidt (Euclidean/flat) measure \cite{szHS} \cite[sec. 14.3]{ingemarkarol}, of generic  two-rebit 
\cite{carl,batle2} and two-qubit density matrices ($\rho$). Here 
$\rho^{PT}$ denotes the partial transpose of the $4 \times 4$ density matrix $\rho$. Nonnegativity of the determinant 
$|\rho^{PT}|$ is both a necessary and sufficient condition for separability in this $2 \times 2$ setting \cite{augusiak}.

In the second set of analyses in \cite{MomentBased}, the {\it univariate} determinantal moments $\left\langle \left\vert \rho^{PT}\right\vert ^{n} \right\rangle$ and $\left\langle \left ( \vert \rho^{PT}\right\vert \left\vert \rho\right\vert)^n
\right\rangle$, induced using the bivariate formulas, served as input to a Legendre-polynomial-based probability distribution reconstruction algorithm of Provost \cite[sec. 2]{Provost}. This yielded estimates of the desired separability probabilities. (The reconstructed probability distributions based on $|\rho^{PT}|$ are defined over the interval $|\rho^{PT}| \in [-\frac{1}{16},\frac{1}{256}]$, while the associated separability probabilities are the cumulative probabilities of these distributions over the nonnegative 
subinterval $|\rho^{PT}| \in [0,\frac{1}{256}]$. We note that for the fully mixed (classical) state, $|\rho^{PT}| = \frac{1}{256}$, while for a maximally entangled state, such as  a Bell state,  $|\rho^{PT}| = -\frac{1}{16}$.)

A highly-intriguing aspect of the (not yet rigorously established) determinantal moment formulas obtained (by C. Dunkl) in \cite[App.D.4]{MomentBased} was that both the two-rebit and two-qubit cases could be encompassed by a 
{\it single} formula, with a Dyson-index-like parameter 
$\alpha$ \cite{MatrixModels} serving to distinguish the two cases. The value $\alpha = \frac{1}{2}$ corresponded to the two-rebit case and $\alpha=1$ to the two-qubit case. Further, the results of the formula for $\alpha=2$ and $n=1$ and 2 have recently been confirmed computationally by Dunkl using the "Moore determinant" (quasideterminant) \cite{Moore,Gelfand} of partially transposed $4 \times 4$ {\it quaternionic} density matrices.

When the probability-distribution-reconstruction algorithm \cite{Provost} was applied 
in \cite{MomentBased} to the two-rebit case ($\alpha=\frac{1}{2}$), employing the first 3,310 moments of $|\rho^{PT}|$, 
a (lower-bound) estimate that was 0.999955 times as large as $\frac{29}{64} \approx 0.453120$ was obtained (cf. \cite[bot. p. 6]{advances}). Analogously, in the two-qubit case ($\alpha =1$), using 2,415 moments, an estimate that was 0.999997066 times as large as $\frac{8}{33} \approx 0.242424$ was derived. This constitutes an appealingly simple rational value that had previously been conjectured in a quite different (non-moment-based) form of analysis, in which "separability functions" had been the main tool employed \cite{slater833}.

Further, the determinantal moment formulas advanced in 
\cite{MomentBased} were then implemented with $\alpha$ set equal to 2. This appears to correspond--as the indicated recent computations of Dunkl support--to the generic 27-dimensional set of quaternionic density matrices \cite{andai,adler}. Quite remarkably, a separability probability estimate, based on 2,325 moments, that was 0.999999987 times as large as $\frac{26}{323} \approx 0.0804954$ was found. 

In this study, we extend these three (independently conducted) moment-based analyses in a more systematic manner, {\it jointly} embracing the 64 integral and half-integral values $\alpha =\frac{1}{2}, 1, \frac{3}{2}, 2,\ldots, 32$. We do this by accelerating, for our specific purposes, the Mathematica probability-distribution-reconstruction program of Provost \cite{Provost},  in a number of ways. Most significantly, we make use of the three-term recurrence relations for the Legendre polynomials. Doing so obviates the need to compute each successive higher-degree Legendre polynomial {\it ab initio}.

In this manner, we were able to obtain--using exact computer arithmetic
throughout--"generalized" separability probability estimates based on 7,501 moments for $\alpha = \frac{1}{2}, 1, \frac{3}{2},\ldots,32$.
Now, the associated two-rebit separability probability estimate increased/improved from 0.999955 times as large as
$\frac{29}{64} \approx 0.453120$ to 0.999989567; the two-qubit separability probability estimate improved from 0.999997066 times as large as
$\frac{8}{33} \approx 0.242424$ to 0.99999986; and the two-quaterbit separability probability  estimate improved from 0.999999987 times as large as $\frac{26}{323} \approx 0.0804954$  to 0.999999999936.

We note that the {\it zeroth}-order approximation (being independent of the particular value of $\alpha$) provided by the Provost probability-distribution-reconstruction algorithm is simply the {\it uniform} distribution over the interval
$[-\frac{1}{16},\frac{1}{256}]$. The corresponding zeroth-order separability probability estimate is the cumulative probability of this distribution over the nonnegative subinterval $[0,\frac{1}{256}]$, that is, $ \frac{1}{256}/(\frac{1}{16} +\frac{1}{256}) =\frac{1}{17} \approx 0.0588235$. So, it certainly appears that speedier convergence of the algorithm occurs for separability probabilities, the true values of which are initially close to 
$\frac{1}{17}$ (such as $\frac{26}{323} \approx 0.0804954$ 
in the quaternionic case).
Convergence also markedly increases as $\alpha$ increases.

It appeared, numerically, that the generalized separability probabilities obtained for all the 32 integral and 32 half-integral values of $\alpha$ employed were rational values 
(not only for the three specific values $\alpha = \frac{1}{2}, 1, 2$ of original focus). With various computational tools and search strategies based upon emerging mathematical properties, we were able to advance additional,  seemingly plausible conjectures as to the exact values for  the remaining 61 values of $\alpha$, as well \cite[p. 6]{SlaterHyper}. 

We fed the sequence of 32 integral-indexed conjectured rational numbers into the FindSequenceFunction command of Mathematica. ("FindSequenceFunction[list] uses earlier elements in list to find candidate simple functions, then validates the functions by looking at later elements. FindSequenceFunction[list] only returns functions that correctly reproduce all elements of list.") This produced a generating formula $P(\alpha)$ incorporating a diversity of hypergeometric functions of the $_{p}F_{p-1}$ type, $p=7,\ldots,11$, {\it all} with argument $z =\frac{27}{64}= (\frac{3}{4})^3$. (We note that $z^{-\frac{1}{2}} = \sqrt{\frac{64}{27}}$ is the "residual entropy for square ice" \cite[p. 412]{finch} (cf. \cite[eqs.[(27), (28)]{ckksr}) \cite{guillera}.) In fact, the Mathematica command succeeds using only the first twenty-eight conjectured rational numbers, but no fewer--so it seems fortunate, our computations were as extensive as they were. 
The formula $P(\alpha)$ produced was quite cumbersome in nature (extending over several pages of output) (cf. \cite[Fig. 3]{SlaterHyper}). 

$P(\alpha)$ did yield  values for {\it half}-integral $\alpha$ (including the 
$\alpha=\frac{1}{2}$ two-rebit conjecture of $\frac{29}{64}$), also exactly fitting our corresponding 32 half-integral   rational-valued conjectured separability probabilities \cite[p. 6]{SlaterHyper}. The process was fully reversible, with the half-integral sequence serving as the inputted one. 

We had been unable to find an equivalent  form of $P(\alpha)$ with fewer than six hypergeometric functions \cite[Fig. 3]{SlaterHyper}. Qing-Hu Hou and colleagues, however,  were able to obtain the remarkably succinct results (\ref{Hou1})-(\ref{Hou3}).
They did so, first observing that the hypergeometric-based formula for $P(\alpha)$ could be expressed as an infinite summation. Letting $P_l(\alpha)$ be the $l$-th such summand, application of Zeilberger's algorithm  \cite{doron} (a method for producing combinatorial identities) (cf. \cite[App. A]{Datta}), then, yielded that 
\begin{equation}
P_l(\alpha) -P_l(\alpha+1) =-P_{l+1}(\alpha) + P_l(\alpha) .
\end{equation}
(The package APCI--available at http://www.combinatorics.net.cn/homepage/hou/--was employed.)
Then, summing over $l$ from 0 to $\infty$, they found that 
\begin{equation}
P(\alpha) -P(\alpha+1)=P_0(\alpha).
\end{equation}
Letting $f(\alpha) =P_0(\alpha)$, the concise summation formula (\ref{Hou1}) is obtained.
(C. Krattenthaler indicated--and Hou agreed--that these results might equally well be derived without recourse to Zeilberger's algorithm.)

We certainly need to indicate, however, that if we do explicitly perform the infinite summation indicated in (\ref{Hou1}), then we revert to a ("nonconcise") form of $P(\alpha)$, again containing six hypergeometric functions. Further, it appears that we can only evaluate (\ref{Hou1}) numerically--but then easily to hundreds and even thousands of digits of precision--giving us extremely high confidence in the specific rational-valued Hilbert-Schmidt separability probabilities advanced.

There remain the important problems of formally verifying the formulas for $P(\alpha)$ (as well as the underlying determinantal moment formulas in \cite{MomentBased}, employed in the probability-distribution reconstruction process), and achieving a better understanding of what these results convey regarding the geometry of quantum states \cite{ingemarkarol}. Further, questions of the asymptotic behavior of the formula ($\alpha \rightarrow \infty$)  and of possible Bures metric 
\cite{szBures,ingemarkarol,slaterJGP,slaterqip,slaterC} counterparts to it, are under investigation \cite{BuresHilbert}.

The foundational paper of {\.Z}yczkowski, Horodecki, Sanpera and Lewenstein,"Volume of the set of separable states" \cite{ZHSL}, did ask for {\it volumes}, not specifically {\it probabilities}. At least, for the two-rebit, two-qubit and two-quaterbit cases, $\alpha =\frac{1}{2}, 1$ and $2$, we can readily convert the corresponding separability probabilities to the separable volumes $\frac{29 \pi ^4}{61931520} =\frac{29 \pi^4}{2^{16} \cdot 3^3 \cdot 5 \cdot 7}$, $\frac{\pi ^6}{449513064000} = \frac{\pi^6}{2^6 \cdot 3^6 \cdot 5^3 \cdot 7^2 \cdot 11^2 \cdot 13}$ and 
$\frac{\pi ^{12}}{3914156909371803494400000} = \frac{\pi^{12}}{2^{14} \cdot 
3^{10} \cdot 5^5 \cdot 7^3 \cdot 11^2 \cdot 13 \cdot 17^2 \cdot 19^2 \cdot 23}$, using the Hilbert-Schmidt volume formulas of \cite[Thms. 1-3]{andai} (cf. \cite{szHS,ingemarkarol}). The determination of separable volumes--as opposed to probabilities--for other values of $\alpha$ than these three appears to be rather problematical, however.

Let us also note that Theorem 2  of \cite{sbz}, in conjunction with the results here, allows us to immediately obtain the separability probabilities of the generic  minimally-degenerate/boundary 8-, 14-, and 26-dimensional two-rebit, two-qubit, and two-quaterbit states, as one-half (that is, $\frac{29}{128}, \frac{4}{33}$ and $\frac{13}{323}$) the separability probabilities of their generic non-degenerate counterparts.

In addition to $P(\alpha)$ assuming rational values for nonnegative integral and half-integral values of $\alpha$, we note that $P(-\frac{1}{2}) = \frac{2}{3}, P(-\frac{1}{4})= 2, P(-1)= \frac{2}{5}$ and $P(-\frac{3}{2})= P(-\frac{1}{2}) =\frac{2}{3}$. A number of other interesting (irrational) values of $P(\alpha)$ are indicated in \cite[p. 9]{SlaterHyper}. For example, $P(-\frac{1}{3}) =2+\frac{3 \Gamma \left(\frac{1}{3}\right)^3}{4 \pi ^2}$, where the term $\frac{3 \Gamma \left(\frac{1}{3}\right)^3}{4 \pi ^2} \approx 1.46099848$ is "Baxter's four-coloring constant" for a triangular lattice \cite[p. 413]{finch}. The first derivative $P'(\alpha)$--similarly to $P(\alpha)$ itself--appears to assume rational values for nonnegative integral and half-integral values of $\alpha$, such as $P'(1) =  -\frac{130577}{457380}$. We further observe that $P'(0) = -2$ and $P''(0) = 40 - 20 \zeta{(2)} = 40 -\frac{10 \pi^2}{3} \approx 7.10132$. 
Thus, let us conclude that the function 
$P(\alpha)$ certainly appears to be an object well worthy of further investigation.

\begin{acknowledgments}
I would like to express appreciation to the Kavli Institute for Theoretical
Physics (KITP)
for computational support in this research, and to Christian Krattenthaler, Charles F. Dunkl, Michael Trott, Jorge Santos, and Qing-Hu Hou and his colleagues for their expert advice.
\end{acknowledgments}

\bibliography{Concise3}

\begin{thebibliography}{27}
\expandafter\ifx\csname natexlab\endcsname\relax\def\natexlab#1{#1}\fi
\expandafter\ifx\csname bibnamefont\endcsname\relax
  \def\bibnamefont#1{#1}\fi
\expandafter\ifx\csname bibfnamefont\endcsname\relax
  \def\bibfnamefont#1{#1}\fi
\expandafter\ifx\csname citenamefont\endcsname\relax
  \def\citenamefont#1{#1}\fi
\expandafter\ifx\csname url\endcsname\relax
  \def\url#1{\texttt{#1}}\fi
\expandafter\ifx\csname urlprefix\endcsname\relax\def\urlprefix{URL }\fi
\providecommand{\bibinfo}[2]{#2}
\providecommand{\eprint}[2][]{\url{#2}}

\bibitem[{\citenamefont{{\.Z}yczkowski
  et~al.}(1998)\citenamefont{{\.Z}yczkowski, Horodecki, Sanpera, and
  Lewenstein}}]{ZHSL}
\bibinfo{author}{\bibfnamefont{K.}~\bibnamefont{{\.Z}yczkowski}},
  \bibinfo{author}{\bibfnamefont{P.}~\bibnamefont{Horodecki}},
  \bibinfo{author}{\bibfnamefont{A.}~\bibnamefont{Sanpera}}, \bibnamefont{and}
  \bibinfo{author}{\bibfnamefont{M.}~\bibnamefont{Lewenstein}},
  \bibinfo{journal}{Phys. Rev. A} \textbf{\bibinfo{volume}{58}},
  \bibinfo{pages}{883} (\bibinfo{year}{1998}).

\bibitem[{\citenamefont{{\.Z}yczkowski and Sommers}(2003)}]{szHS}
\bibinfo{author}{\bibfnamefont{K.}~\bibnamefont{{\.Z}yczkowski}}
  \bibnamefont{and} \bibinfo{author}{\bibfnamefont{H.-J.}
  \bibnamefont{Sommers}}, \bibinfo{journal}{J. Phys. A}
  \textbf{\bibinfo{volume}{36}}, \bibinfo{pages}{10115} (\bibinfo{year}{2003}).

\bibitem[{\citenamefont{Bengtsson and {\.Z}yczkowski}(2006)}]{ingemarkarol}
\bibinfo{author}{\bibfnamefont{I.}~\bibnamefont{Bengtsson}} \bibnamefont{and}
  \bibinfo{author}{\bibfnamefont{K.}~\bibnamefont{{\.Z}yczkowski}},
  \emph{\bibinfo{title}{Geometry of Quantum States}}
  (\bibinfo{publisher}{Cambridge}, \bibinfo{address}{Cambridge},
  \bibinfo{year}{2006}).

\bibitem[{\citenamefont{Batle et~al.}(2003)\citenamefont{Batle, Plastino,
  Casas, and Plastino}}]{batle2}
\bibinfo{author}{\bibfnamefont{J.}~\bibnamefont{Batle}},
  \bibinfo{author}{\bibfnamefont{A.~R.} \bibnamefont{Plastino}},
  \bibinfo{author}{\bibfnamefont{M.}~\bibnamefont{Casas}}, \bibnamefont{and}
  \bibinfo{author}{\bibfnamefont{A.}~\bibnamefont{Plastino}},
  \bibinfo{journal}{Opt. Spect.} \textbf{\bibinfo{volume}{94}},
  \bibinfo{pages}{759} (\bibinfo{year}{2003}).

\bibitem[{\citenamefont{Slater and Dunkl}(2012)}]{MomentBased}
\bibinfo{author}{\bibfnamefont{P.~B.} \bibnamefont{Slater}} \bibnamefont{and}
  \bibinfo{author}{\bibfnamefont{C.~F.} \bibnamefont{Dunkl}},
  \bibinfo{journal}{J. Phys. A} \textbf{\bibinfo{volume}{45}},
  \bibinfo{pages}{095305} (\bibinfo{year}{2012}).

\bibitem[{\citenamefont{Caves et~al.}(2001)\citenamefont{Caves, Fuchs, and
  Rungta}}]{carl}
\bibinfo{author}{\bibfnamefont{C.~M.} \bibnamefont{Caves}},
  \bibinfo{author}{\bibfnamefont{C.~A.} \bibnamefont{Fuchs}}, \bibnamefont{and}
  \bibinfo{author}{\bibfnamefont{P.}~\bibnamefont{Rungta}},
  \bibinfo{journal}{Found. Phys. Letts.} \textbf{\bibinfo{volume}{14}},
  \bibinfo{pages}{199} (\bibinfo{year}{2001}).

\bibitem[{\citenamefont{Augusiak et~al.}(2008)\citenamefont{Augusiak,
  Horodecki, and Demianowicz}}]{augusiak}
\bibinfo{author}{\bibfnamefont{R.}~\bibnamefont{Augusiak}},
  \bibinfo{author}{\bibfnamefont{R.}~\bibnamefont{Horodecki}},
  \bibnamefont{and}
  \bibinfo{author}{\bibfnamefont{M.}~\bibnamefont{Demianowicz}},
  \bibinfo{journal}{Phys. Rev.} \textbf{\bibinfo{volume}{77}},
  \bibinfo{pages}{030301(R)} (\bibinfo{year}{2008}).

\bibitem[{\citenamefont{Provost}(2005)}]{Provost}
\bibinfo{author}{\bibfnamefont{S.~B.} \bibnamefont{Provost}},
  \bibinfo{journal}{Mathematica J.} \textbf{\bibinfo{volume}{9}},
  \bibinfo{pages}{727} (\bibinfo{year}{2005}).

\bibitem[{\citenamefont{Dumitriu and Edelman}(2002)}]{MatrixModels}
\bibinfo{author}{\bibfnamefont{I.}~\bibnamefont{Dumitriu}} \bibnamefont{and}
  \bibinfo{author}{\bibfnamefont{A.}~\bibnamefont{Edelman}},
  \bibinfo{journal}{J. Math. Phys.} \textbf{\bibinfo{volume}{43}},
  \bibinfo{pages}{5830} (\bibinfo{year}{2002}).

\bibitem[{\citenamefont{Moore}(1922)}]{Moore}
\bibinfo{author}{\bibfnamefont{E.~H.} \bibnamefont{Moore}},
  \bibinfo{journal}{Bull. Amer. Math. Soc.} \textbf{\bibinfo{volume}{28}},
  \bibinfo{pages}{161} (\bibinfo{year}{1922}).

\bibitem[{\citenamefont{Gelfand et~al.}(2005)\citenamefont{Gelfand, Gelfand,
  Retakh, and Wilson}}]{Gelfand}
\bibinfo{author}{\bibfnamefont{I.}~\bibnamefont{Gelfand}},
  \bibinfo{author}{\bibfnamefont{S.}~\bibnamefont{Gelfand}},
  \bibinfo{author}{\bibfnamefont{V.}~\bibnamefont{Retakh}}, \bibnamefont{and}
  \bibinfo{author}{\bibfnamefont{R.~L.} \bibnamefont{Wilson}},
  \bibinfo{journal}{Adv. Math.} \textbf{\bibinfo{volume}{193}},
  \bibinfo{pages}{56} (\bibinfo{year}{2005}).

\bibitem[{\citenamefont{Slater}(2010)}]{advances}
\bibinfo{author}{\bibfnamefont{P.~B.} \bibnamefont{Slater}},
  \bibinfo{journal}{J. Phys. A} \textbf{\bibinfo{volume}{43}},
  \bibinfo{pages}{195302} (\bibinfo{year}{2010}).

\bibitem[{\citenamefont{Slater}(2007)}]{slater833}
\bibinfo{author}{\bibfnamefont{P.~B.} \bibnamefont{Slater}},
  \bibinfo{journal}{J. Phys. A} \textbf{\bibinfo{volume}{40}},
  \bibinfo{pages}{14279} (\bibinfo{year}{2007}).

\bibitem[{\citenamefont{Andai}(2006)}]{andai}
\bibinfo{author}{\bibfnamefont{A.}~\bibnamefont{Andai}}, \bibinfo{journal}{J.
  Phys. A} \textbf{\bibinfo{volume}{39}}, \bibinfo{pages}{13641}
  (\bibinfo{year}{2006}).

\bibitem[{\citenamefont{Adler}(1995)}]{adler}
\bibinfo{author}{\bibfnamefont{S.~L.} \bibnamefont{Adler}},
  \emph{\bibinfo{title}{Quaternionic quantum mechanics and quantum fields}}
  (\bibinfo{publisher}{Oxford}, \bibinfo{address}{New York},
  \bibinfo{year}{1995}).

\bibitem[{\citenamefont{Slater}({\natexlab{a}})}]{SlaterHyper}
\bibinfo{author}{\bibfnamefont{P.~B.} \bibnamefont{Slater}},
  \eprint{arXiv:1203.4498}.

\bibitem[{\citenamefont{Finch}(2003)}]{finch}
\bibinfo{author}{\bibfnamefont{S.~R.} \bibnamefont{Finch}},
  \emph{\bibinfo{title}{Mathematical Constants}}
  (\bibinfo{publisher}{Cambridge}, \bibinfo{address}{New York},
  \bibinfo{year}{2003}).

\bibitem[{\citenamefont{Krattenthaler and Rao}(2005)}]{ckksr}
\bibinfo{author}{\bibfnamefont{C.}~\bibnamefont{Krattenthaler}}
  \bibnamefont{and} \bibinfo{author}{\bibfnamefont{K.~S.} \bibnamefont{Rao}},
  \bibinfo{journal}{Symmetries in Science} \textbf{\bibinfo{volume}{XI}},
  \bibinfo{pages}{355} (\bibinfo{year}{2005}).

\bibitem[{\citenamefont{Guillera}(2011)}]{guillera}
\bibinfo{author}{\bibfnamefont{J.}~\bibnamefont{Guillera}},
  \bibinfo{journal}{Ramanujan J.} \textbf{\bibinfo{volume}{26}},
  \bibinfo{pages}{369} (\bibinfo{year}{2011}).

\bibitem[{\citenamefont{Zeilberger}(1990)}]{doron}
\bibinfo{author}{\bibfnamefont{D.}~\bibnamefont{Zeilberger}},
  \bibinfo{journal}{Discr. Math.} \textbf{\bibinfo{volume}{80}},
  \bibinfo{pages}{207} (\bibinfo{year}{1990}).

\bibitem[{\citenamefont{Datta}(2010)}]{Datta}
\bibinfo{author}{\bibfnamefont{A.}~\bibnamefont{Datta}},
  \bibinfo{journal}{Phys. Rev. A} \textbf{\bibinfo{volume}{81}},
  \bibinfo{pages}{052312} (\bibinfo{year}{2010}).

\bibitem[{\citenamefont{Sommers and {\.Z}yczkowski}(2003)}]{szBures}
\bibinfo{author}{\bibfnamefont{H.-J.} \bibnamefont{Sommers}} \bibnamefont{and}
  \bibinfo{author}{\bibfnamefont{K.}~\bibnamefont{{\.Z}yczkowski}},
  \bibinfo{journal}{J. Phys. A} \textbf{\bibinfo{volume}{36}},
  \bibinfo{pages}{10083} (\bibinfo{year}{2003}).

\bibitem[{\citenamefont{Slater}(2005)}]{slaterJGP}
\bibinfo{author}{\bibfnamefont{P.~B.} \bibnamefont{Slater}},
  \bibinfo{journal}{J. Geom. Phys.} \textbf{\bibinfo{volume}{53}},
  \bibinfo{pages}{74} (\bibinfo{year}{2005}).

\bibitem[{\citenamefont{Slater}(2002)}]{slaterqip}
\bibinfo{author}{\bibfnamefont{P.~B.} \bibnamefont{Slater}},
  \bibinfo{journal}{Quant. Info. Proc.} \textbf{\bibinfo{volume}{1}},
  \bibinfo{pages}{397} (\bibinfo{year}{2002}).

\bibitem[{\citenamefont{Slater}(2000)}]{slaterC}
\bibinfo{author}{\bibfnamefont{P.~B.} \bibnamefont{Slater}},
  \bibinfo{journal}{Euro. Phys. J. B} \textbf{\bibinfo{volume}{17}},
  \bibinfo{pages}{471} (\bibinfo{year}{2000}).

\bibitem[{\citenamefont{Slater}({\natexlab{b}})}]{BuresHilbert}
\bibinfo{author}{\bibfnamefont{P.~B.} \bibnamefont{Slater}},
  \eprint{arXiv:1207.1297}.

\bibitem[{\citenamefont{Szarek et~al.}(2006)\citenamefont{Szarek, Bengtsson,
  and {\.Z}yczkowski}}]{sbz}
\bibinfo{author}{\bibfnamefont{S.}~\bibnamefont{Szarek}},
  \bibinfo{author}{\bibfnamefont{I.}~\bibnamefont{Bengtsson}},
  \bibnamefont{and}
  \bibinfo{author}{\bibfnamefont{K.}~\bibnamefont{{\.Z}yczkowski}},
  \bibinfo{journal}{J. Phys. A} \textbf{\bibinfo{volume}{39}},
  \bibinfo{pages}{L119} (\bibinfo{year}{2006}).

\end{thebibliography}

\end{document}